\documentclass[a4paper,11pt]{article}
\pdfoutput=1
\usepackage{jheppub} 
 \usepackage{ulem}
\usepackage[lofdepth,lotdepth,caption=false]{subfig}
\usepackage[T1]{fontenc}

\usepackage{amsmath}
\usepackage{enumerate}
\usepackage{slashed}

\title{Constraining the monochromatic gamma-rays from dark matter annihilation by the LHC}

\author[a]{Arman Esmaili,}
\author[b]{Sara Khatibi}
\author[b]{and Mojtaba Mohammadi Najafabadi}

\affiliation[a]{Departamento de F\'isica, Pontif\'icia Universidade Cat\'olica do Rio de Janeiro, C.~P.~38071, 22452-970, Rio de Janeiro, Brazil}
\affiliation[b]{School of Particles and Accelerators, Institute for Research in Fundamental Sciences~(IPM) P.O. Box 19395-5531, Tehran, Iran}

\emailAdd{arman@puc-rio.br}
\emailAdd{s.khatibi@ipm.ir}
\emailAdd{mojtaba.mohammadi.najafabadi@cern.ch}

\abstract{The installation of forward detectors in CMS and ATLAS turn the LHC to an effective photon-photon collider. The elastic scattering of the beam-protons via the emission of photons, which can be identified by tagging the intact protons in the forward detectors, provides a powerful diagnostic of the central production of new particles through photon-photon annihilation. In this letter we study the central production of dark matter particles and the potential of LHC to constrain the cross section of this process. By virtue of the crossing symmetry, this limit can immediately be used to constrain the production of monochromatic gamma-rays in dark matter annihilation, a smoking gun signal under investigation in indirect dark matter searches. We show that with the integrated luminosity $\mathcal{L}=30~{\rm fb}^{-1}$ in LHC at center-of-mass energy $\sqrt{s} = 13$~TeV, for dark matter masses $\sim (50-600)$~GeV, a model-independent constraint on the cross section of dark matter annihilation to monochromatic gamma-rays at the same order of magnitude as the current Fermi-LAT and the future limits from CTA, can be obtained.}

\begin{document}
\maketitle
\flushbottom

%%%%%%%%%%%%%%%%%%%%%%%%%%
%%%%%%%%%%%%%%%%%%%%%%%%%%
\section{Introduction} \label{sec:intro}
%%%%%%%%%%%%%%%%%%%%%%%%%%
%%%%%%%%%%%%%%%%%%%%%%%%%%

A plethora of evidence, all from gravitational interactions, ranging from galactic to cosmic scales, put forward the existence of a new particle(s) responsible for the missing mass in the Universe, coined Dark Matter (DM). Although this new particle(s) cannot be accommodated within the field content of the Standard Model (SM), in most of the models a weak interaction with the SM particles is assumed (typically via exchange of particles mediating the force between the standard and dark sectors). The weak scale of this interaction is especially motivated in the Weakly Interacting Massive Particle (WIMP) scenario, where the right abundance of DM particles in the present time can be obtained by the freeze-out mechanism of DM production in the early universe, with DM particle masses from $\sim$ GeV to few hundreds of TeV. This assumed interaction, generally described by higher dimensional terms in the Lagrangian within an effective field theory approach, leads to various strategies in looking for the elusive DM particle, commonly categorized as direct, indirect and collider searches. In the direct detection the experimental signature of DM detection is the recoil of nucleus in the scattering off the DM particles; while the underlying process in the indirect and collider searches is DM annihilation/decay and creation, respectively. 

In the indirect DM searches, experimental signature is the excess (over the relevant background) of the stable particles in the products of DM annihilation/decay. Among the stable particles (usually $\gamma$, $\nu$, $e^+$, $\bar{p}$ and anti-deuteron), the $\gamma$ rays are one of the promising messengers in the search for DM. Generally, the $\gamma$ rays from DM annihilation/decay can be produced in three different ways: {\it i)} from the radiation and hadronization of annihilation/decay products, which lead to a continuous spectrum from $m_{\rm DM}$ ($m_{\rm DM}/2$ for the decaying DM) to lower energies (in fact, larger spectrum in the lower energies); {\it ii)} from the radiative processes of the annihilation/decay products (such as synchrotron radiation or the inverse-Compton scattering), which lead to a continuous spectrum down to very low energies~\cite{Colafrancesco:2006he,Colafrancesco:2005ji}; {\it iii)} from the annihilation/decay of DM particles either directly to $\gamma\gamma$, $\gamma Z$ and $\gamma h$ states (through loops)~\cite{Bergstrom:1997fj,Ullio:1997ke,Bergstrom:2004nr,Gustafsson:2007pc,Bergstrom:1988fp,Bergstrom:1997fh,Bern:1997ng,Ibarra:2014vya,Jackson:2013pjq,Duerr:2015wfa,Garcia-Cely:2016hsk}, or through intermediate states leading to narrow box-shaped spectral lines~\cite{Ibarra:2012dw,Ibarra:2013eda}, or internal bremsstrahlung~\cite{Bergstrom:2004cy,Beacom:2004pe,Barger:2011jg,Bringmann:2007nk}, which lead to a (almost) monochromatic line in the spectrum. The main challenge in the indirect DM searches by gamma rays is discrimination of the speculated DM signal from the ubiquitous continuous featureless background spectrum from astrophysical objects, which leave the monochromatic line searches a promising approach.

The collider searches for DM are essentially based on the inverse of the underlying process in indirect searches; {\it i.e.}, the creation of DM particles in the scattering of SM particles. Although the main channel of DM production at the LHC is $pp\rightarrow\chi\bar{\chi}$, in practice looking for such reaction (with the signature of a large Missing Transverse Energy, MET) is hopeless since the initial interaction of protons ({\it i.e.}, whether it happened or not) cannot be tagged in this reaction. Consequently, the conventional channels of DM searches at the LHC, although higher order processes in perturbation, are the DM plus a hadronic/weak production, such as mono-jet~\cite{CMS:2016pod,Sun:2014ppa}, mono-photon~\cite{CMS:2016fnh} and single-$Z$~\cite{CMS:2016hmx,Aad:2014vka,Aaboud:2016qgg} searches. However, there is a configuration that this limitation can be elevated thanks to the already implemented forward detectors at CMS, the CMS-TOTEM Precision Proton Spectrometer (CT-PPS)~\cite{pps}, and the planned detectors at ATLAS, the ATLAS Forward Physics (AFP)~\cite{afp}. In this paper we consider the elastic scattering of the protons via photon-photon fusion\footnote{The central diffractive processes where one of the protons dissociate will be left for future studies.} which corresponds to the following process: the two incoming protons at the LHC emit photons and remain intact, except of course loosing some energy. The outgoing protons (almost collinear) can be tagged in the forward detectors with a high-precision measurement of their energies and transverse momenta, which signal the occurrence of elastic scattering via emission of photons and provides the energy of the emitted photons (of course at the statistical level). The cross section of the whole process can be factorized in the Equivalent Photon Approximation (EPA) \cite{Budnev:1974de} to the probability of photon emission (at an specific energy) from each of the protons convoluted with the cross section of photon-photon fusion. 

The photon-photon fusion has been already studied in the context of Central Exclusive Production (CEP) of Higgs, leptons, beyond SM (such as SUSY) particles, etc~\cite{Fichet:2016pvq,Fichet:2015vvy,Fichet:2016clq,sp1,Harland-Lang:2016qjy,deFavereaudeJeneret:2009db,Sahin:2010zr,Fayazbakhsh:2015xba}. In this paper we consider the fusion process $\gamma\gamma\rightarrow\chi\bar{\chi}$, where the experimental signature consist of two intact protons in the forward detectors plus MET, that is nothing in the central detector. The same signature in the context of invisible decays of Higgs has been considered in~\cite{Belotsky:2004ex}. The fusion $\gamma\gamma\rightarrow\chi\bar{\chi}$ is the inverse of the process leading to spectrum ({\it iii}) discussed above (DM annihilation to monochromatic gamma rays) and searched for in the indirect DM detection. Due to the general principle of detailed balance (or crossing symmetry), the cross section of this fusion is equal to the DM annihilation cross section $\chi\bar{\chi}\rightarrow\gamma\gamma$ and so the LHC can indirectly contribute to the gamma line searches from DM annihilation. We will discuss this synergy in detail and calculate the sensitivity of the LHC to the cross section of DM annihilation to gamma rays $\sigma(\chi\bar{\chi}\rightarrow\gamma\gamma)$. We will show that with the forward detectors in CMS and ATLAS, it is possible to probe the parameter space of the dark matter models (with dark matter mass $\sim(50-600)$~GeV) that predict ${\rm Br}(\chi\bar{\chi}\rightarrow\gamma\gamma)=\langle\sigma(\chi\bar{\chi}\rightarrow\gamma\gamma) v\rangle/\langle\sigma v\rangle \gtrsim \left(10^{-3}-10^{-2}\right)$, where $\langle\sigma v\rangle$ is the total annihilation cross section with the value $3\times10^{-26}~{\rm cm}^3/{\rm s}$ to satisfy the requirement of thermal freeze-out mechanism. Achieving this limit requires a robust rejection of background events that will be discussed in detail.   

The paper is organized as follows: in section~\ref{sec:cep} we calculate the flux of emitted photons from protons. Sections~\ref{sec:SMbkg}, \ref{sec:pile} and \ref{sec:brem} are devoted to the detailed estimation of the background events that play an important role in our analysis. The sensitivity of the LHC to $\langle\sigma(\chi\bar{\chi}\rightarrow\gamma\gamma) v\rangle$ and its comparison with the current and future limits is discussed in section~\ref{sec:sen}. The conclusions are provided in section~\ref{sec:Conclusions}.

%%%%%%%%%%%%%%%%%%%%%%%%%
%%%%%%%%%%%%%%%%%%%%%%%%%
\section{Generalities on the photon-photon fusion at the LHC} 
\label{sec:cep}
%%%%%%%%%%%%%%%%%%%%%%%%%
%%%%%%%%%%%%%%%%%%%%%%%%%
 
The CEP processes in high energy particle colliders provide a very clean environment in the search for SM and beyond the SM physics. At the LHC, the {\it conventional} CEP is the class of reactions 
\begin{eqnarray}
p+p \rightarrow  p+X +p,
\end{eqnarray}
where the colliding protons emerge intact and are observed by the forward detectors, while the produced state $X$ is fully measured by the central detectors. The requirement of intactness of the two protons to be tagged in the forward detectors leave the following cases among the possible scenarios for $pp$ scattering: either both the protons emit a photon, the so-called (doubly) elastic CEP, or one (or both) of the protons emit a color-singlet state, called the diffractive scattering; where the former is the reaction of interest in this paper. The emitted photons, which alter the energy and direction of the protons (in the low photon energy limit, the scattering angle is small), fuse and produce the state $X$. In the conventional CEP process, where the state $X$ (either a SM or beyond SM particle) decay, there is a large rapidity gap between the intact protons and the centrally produced particles, i.e. $\Delta \eta \gtrsim 3$ \cite{Albrow:2010yb} where $\eta=-\ln\left(\tan\left(\frac{\theta}{2}\right)\right)$ and $\theta$ is the angle between particle's momentum and beam direction. In this case the central detectors measure the low rapidity products in the decay of $X$ and the forward detectors tag the protons in high rapidity range and measure their momenta, where the latter provides the invariant mass of the central state $X$. The forward detectors can observe the intact protons in an interval $\xi_{\rm min}< \xi < \xi_{\rm max}$ (which is called the forward detector acceptance region) where
\begin{equation}
\xi \equiv \frac{E_{\rm loss}}{E_{p}}=\frac{E_{p}-E_{p^\prime}}{E_{p}}~,
\end{equation}
where $E_{p}$ and $E_{p^\prime}$ are the energies of the incoming and scattered proton, respectively. For the CT-PPS (AFP) the values are $\xi_{\rm min}=0.0015$ (0.0015) and $\xi_{\rm max}=0.5$ (0.15) \cite{Albrow:2008pn, cmstotem, Royon:2014kta}.

The CEP process can be factorized into the convolution of the following processes: photons emission from protons the photon-photon fusion producing the state $X$:
\begin{eqnarray}\label{eq:cep}
pp\rightarrow p + \gamma\gamma + p \nonumber \\
\text{followed by:}~\gamma\gamma \rightarrow  X~.
\end{eqnarray} 
The double differential energy spectrum of photons emitted from a proton with energy $E_p$, in the EPA approximation~\cite{Budnev:1974de,Terazawa:1973tb}, is given by
\begin{eqnarray}
\frac{{\rm d}^2N}{{\rm d}E_{\gamma}{\rm d}Q^2} = \frac{\alpha}{\pi E_\gamma Q^{2}} \left[\left(1-\frac{E_{\gamma}}{E_p}\right)\left(1-\frac{Q^{2}_{\rm min}}{Q^{2}}\right)F_{E}+\frac{E^{2}_{\gamma}}{2E_p^{2}}F_{M}\right]~,
\end{eqnarray}
where $\alpha$ is the fine-structure constant and $Q^2$ is the photon's virtuality, which is also equal to the transverse momentum of the proton after the photon emission, $p_{\rm T}$; that is $Q^2\simeq p_{\rm T}^2$. The kinematically allowed minimum value of photon's virtuality is $Q^2_{\rm min,kin} = (m_p E_\gamma)^2/(E_p^2-E_pE_\gamma)$ where $m_p$ is the proton's mass. The $F_E$ and $F_M$ are functions of the electric, $G_E$, and magnetic, $G_M$, form factors of the proton
\begin{equation}
F_E = \frac{4m_p^2G^2_E+Q^2G_M^2}{4m_p^2+Q^2}\qquad , \qquad F_M = G_M^2~,
\end{equation}
where in the dipole approximation~\cite{ee} are given by
\begin{equation}
G_M^2 = \mu_p^2G_E^2 = \mu_p^2\left(1+\frac{Q^2}{Q_0^2}\right)^{-4}~,
\end{equation}    
with $Q_0^2=0.71~{\rm GeV}^2$ and $\mu_p^2=7.78$.

\begin{figure}[t!]
\centering
\subfloat[]{
\includegraphics[width=0.5\textwidth]{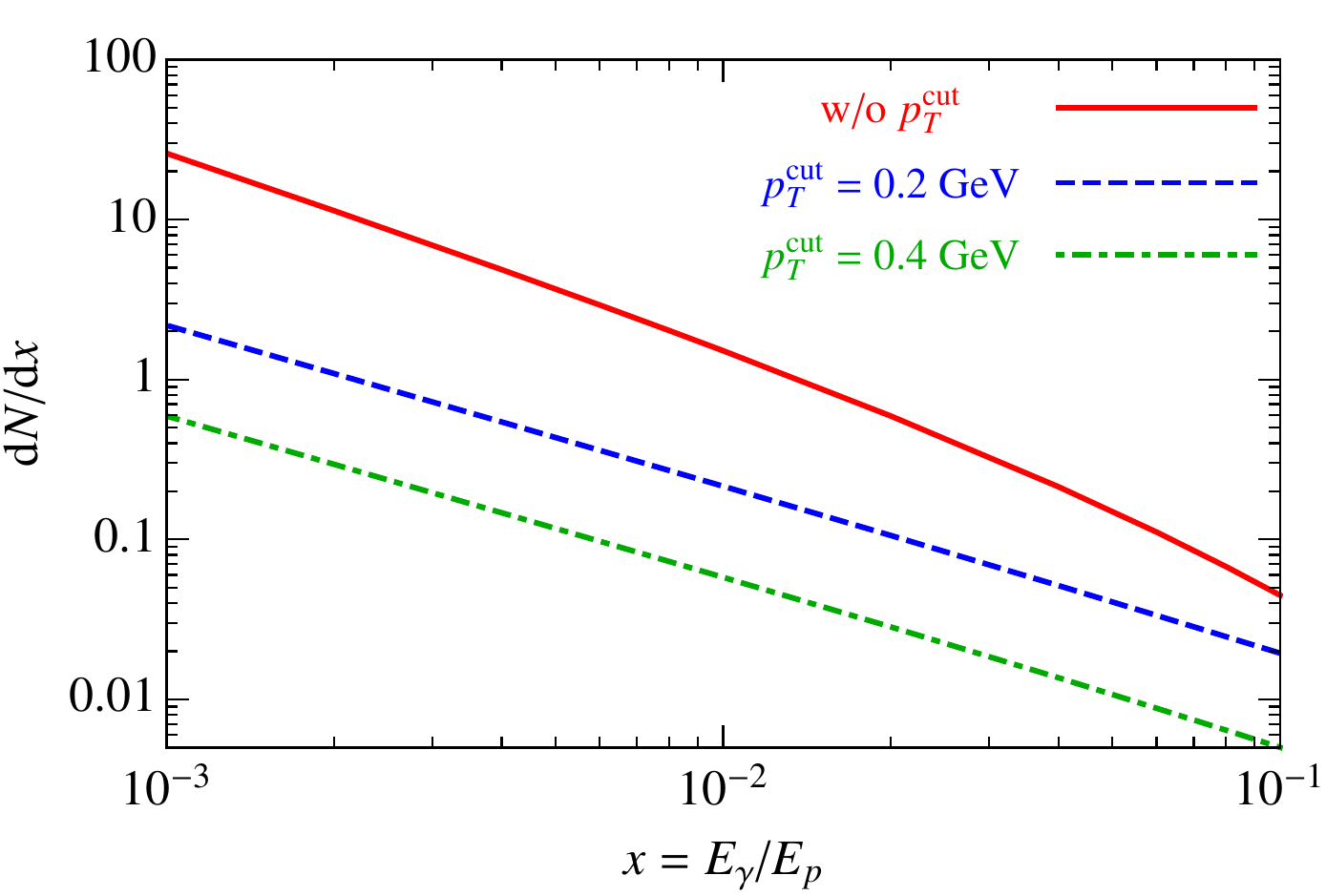}
\label{fig:dndx}
}
\subfloat[]{
\includegraphics[width=0.5\textwidth]{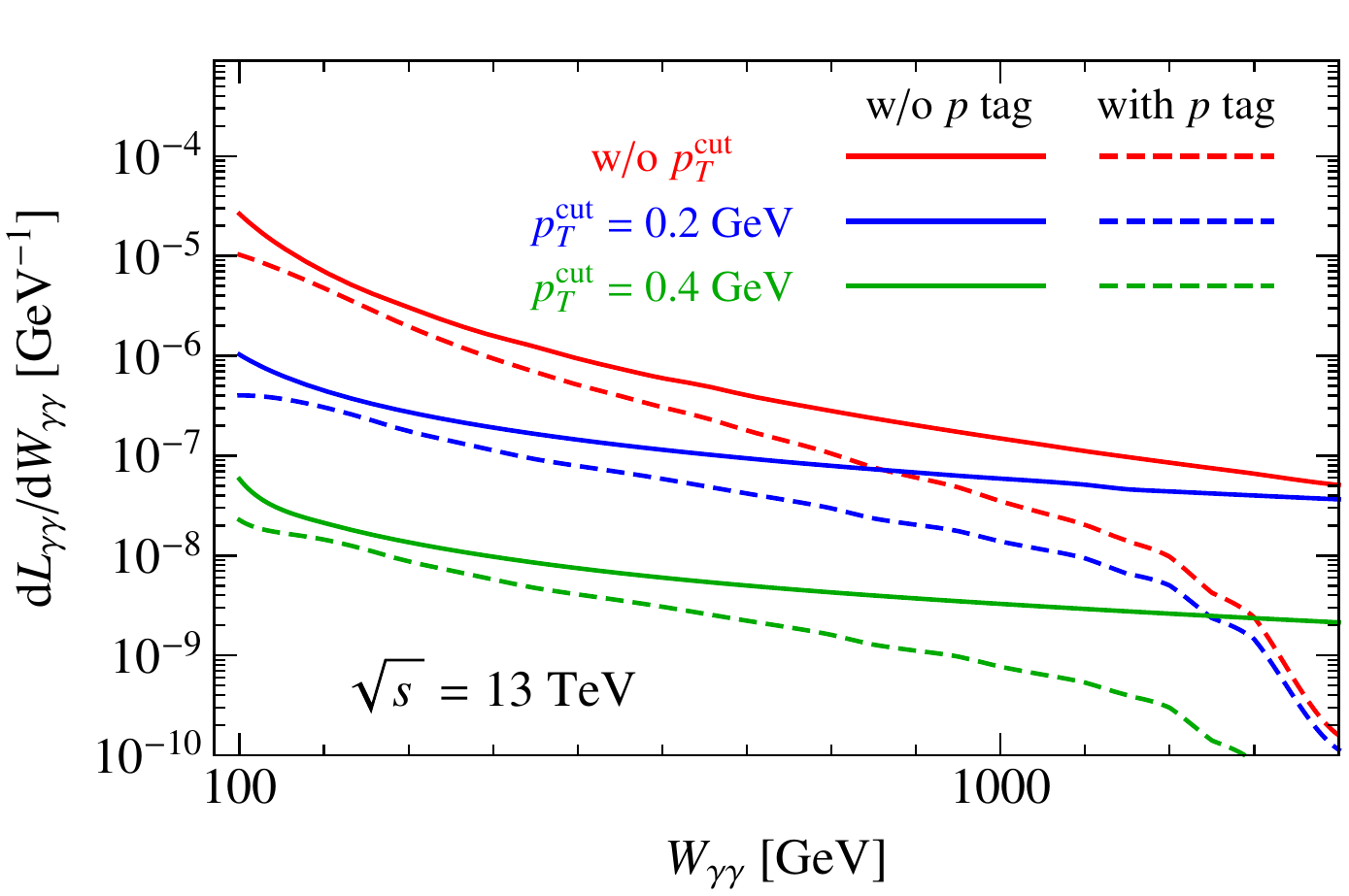}
\label{fig:lum}
}
\caption{\label{fig:spec} a) The spectrum of photons (log-log plot) as function of $x=E_{\gamma}/E_p$. b) Photon-photon luminosity (log-linear plot) as function of center of mass energy $W_{\gamma\gamma}$. The color codes are explained tin the main text.}
\end{figure}

The total cross section of the CEP process in Eq.~(\ref{eq:cep}) can be written as the following convolution:
\begin{equation}
\sigma = \int \sigma_{\gamma \gamma \rightarrow X}(W_{\gamma\gamma})~\frac{{\rm d}L_{\gamma \gamma}}{{\rm d}W_{\gamma\gamma}}(W_{\gamma\gamma})~{\rm d}W_{\gamma\gamma}~,
\end{equation}
where $\sigma_{\gamma \gamma \rightarrow X}$ is the cross section of producing state $X$ in the annihilation of two photons with center of mass energy $W_{\gamma\gamma}$ and ${\rm d}L_{\gamma \gamma}/{\rm d}W_{\gamma\gamma}$ is the luminosity function of the two photons emission. The luminosity function ${\rm d}L_{\gamma \gamma}/{\rm d}W_{\gamma\gamma}$ can be calculated by integrating photon spectra from both protons, $f(E_{\gamma_1})f(E_{\gamma_2})$, over the photon energies and keeping the two-photon invariant mass fixed to $W_{\gamma\gamma}$. The photon spectrum can be obtained by
\begin{eqnarray}\label{eq:spec}
f(E_{\gamma}) = \int_{Q^{2}_{\rm min}}~\frac{{\rm d}^2N}{{\rm d}E_{\gamma}{\rm d}Q^2}~{\rm d}Q^2~.
\end{eqnarray}
Setting $Q^{2}_{\rm min}=Q^{2}_{\rm min,kin}$, the above integration gives the spectrum of photons emitted from a proton. As we will see in the next section, in order to reject some backgrounds, we are interested in the spectrum of photons corresponding to a cut on the transverse momentum of protons, $p_{\rm T}^{\rm cut}$; that is the spectrum of photons where the forward protons have $p_{\rm T}$ larger than $p_{\rm T}^{\rm cut}$. This spectrum can be obtained from Eq.~(\ref{eq:spec}) by setting $Q^{2}_{\rm min}=(p_{\rm T}^{\rm cut})^2$. The upper limit of the integration in Eq.~(\ref{eq:spec}) can safely be set to $\simeq 2~{\rm GeV}^2$ since the contribution of larger $Q^2$ values is negligible. The photon spectrum $f(E_{\gamma})$ decreases rapidly by increasing the photon energy $E_\gamma$, as can be seen in Figure~\ref{fig:dndx} that shows the normalized spectrum of photons ${\rm d}N/{\rm d}x\equiv E_pf(E_\gamma)$ as function of $x=E_\gamma/E_p$. In Figure~\ref{fig:dndx} the red solid curve is the photon spectrum without any cut on the transverse momenta of photons, that is setting $Q^{2}_{\rm min}=Q^{2}_{\rm min,kin}$ in Eq.~(\ref{eq:spec}). The blue (dashed) and green (dot-dashed) curves correspond to the cuts $p_{\rm T}^{\rm cut}=0.2~{\rm GeV}$ and $0.4~{\rm GeV}$, respectively. As can be seen, applying the cut $p_{\rm T}^{\rm cut}=0.2~{\rm GeV}$ on the transverse momenta of protons leads to an order of magnitude reduction in the photon spectrum at $x\simeq10^{-3}$. The drop in the spectrum is smaller for higher values of $x$. The photon luminosity also dominates at low invariant masses $W_{\gamma\gamma}$ ($=2\sqrt{E_{\gamma_1}E_{\gamma_2}}$), which can be seen in Figure~\ref{fig:lum} that shows the {\it relative} luminosity (to the luminosity of protons in the LHC) of two-photon emission for proton beam of energy $E_p=6.5$~TeV. In Figure~\ref{fig:lum} the red, blue and green solid curves show the ${\rm d}L_{\gamma \gamma}/{\rm d}W_{\gamma\gamma}$ respectively for ``no $p_{\rm T}^{\rm cut}$'', $p_{\rm T}^{\rm cut}=0.2~{\rm GeV}$ and $p_{\rm T}^{\rm cut}=0.4~{\rm GeV}$ cases. Application of the cut $p_{\rm T}^{\rm cut}=0.2~{\rm GeV}$ leads to $\sim$ one order of magnitude drop in the photon luminosity for the low invariant masses ($\sim100$~GeV), while the reduction is smaller for the larger $W_{\gamma\gamma}$. The dashed curves in Figure~\ref{fig:lum} show the {\it effective} luminosity of photons after taking into account the acceptance and efficiency of forward detectors in the tagging of forward detectors. The efficiency of forward detectors decrease by the increase in the energy of emitted photons, which leads to strong reduction of effective luminosity for large values of $W_{\gamma\gamma}$. The total (integrated) luminosity in a range of $W_{\gamma\gamma}$ can be obtained by integrating ${\rm d}L_{\gamma \gamma}/{\rm d}W_{\gamma\gamma}$. Evidently (see Figure~\ref{fig:lum}) the total luminosity of photons with invariant mass larger than $W_0$, $L_{\gamma\gamma}(W_{\gamma\gamma}>W_0)$, decreases by increasing the $W_0$, which consequently leads to a reduction in the cross section for production of the state $X$ at large invariant masses. 

%%%%%%%%%            figure 1           %%%%%%%%%%%%%
%%%%%%%%%%%%%%%%%%%%%%%%%%%%%%%%%
%%%%%%%%%%%%%%%%%%%%%%%%%%%%%%%
%%%%%%%%%%%%%%%%%%%%%%%%%%%%%%%

%%%%%%%%%%%%%%%%%%%%%%%%%%%%%%
%%%%%%%%%%%%%%%%%%%%%%%%%%%%%%
\section{Constraining the $\chi\bar{\chi}\rightarrow\gamma\gamma$ by the LHC}
\label{sec:analysis}
%%%%%%%%%%%%%%%%%%%%%%%%%%%%%%
%%%%%%%%%%%%%%%%%%%%%%%%%%%%%%

\begin{figure}[t!]
\centering
{\includegraphics[width = 3in]{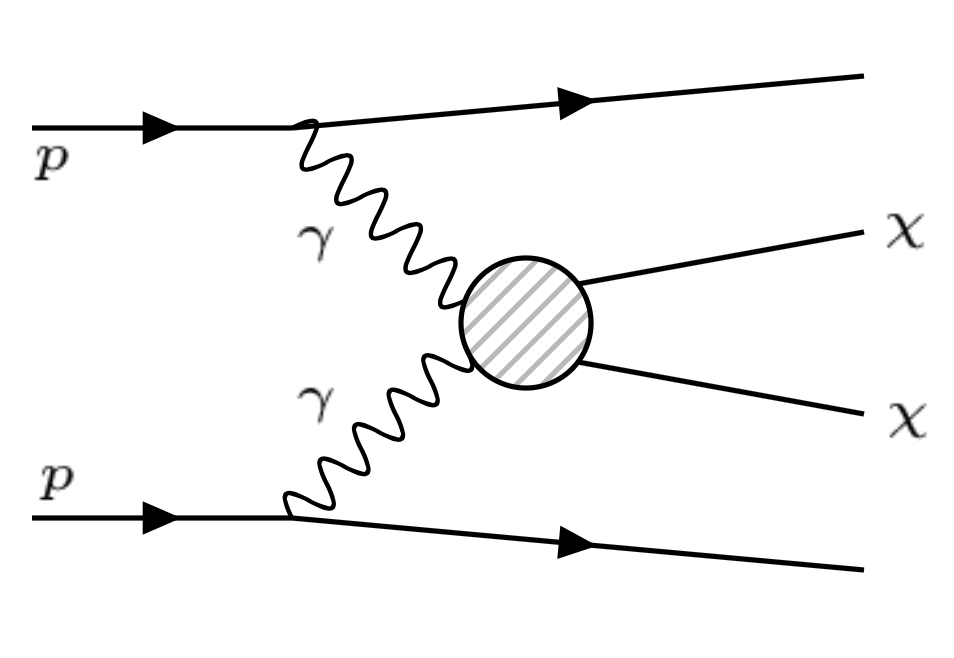}} 
\caption{The Feynman diagram for $p p \rightarrow p + \gamma \gamma + p \rightarrow p + \chi\bar{\chi} +p$ at the LHC. }
\label{fig:fdiagram}
\end{figure}

The process of interest in this paper is the central production of DM particles through the photon-photon fusion with a representative Feynman digram shown in Figure~\ref{fig:fdiagram}. The signature includes two intact protons to be tagged in the forward detectors plus a large missing energy (that is basically no recorded activity in the central detector); {\it i.e.}, $p p \rightarrow p + \gamma \gamma + p \rightarrow p p +  \slashed{E}$ where $\slashed{E}$ denotes the missing energy. The precise energy and transverse momentum measurements in the forward detectors provide the invariant mass of the two-photon system, $W_{\gamma\gamma}$, which through the reaction $\gamma\gamma\rightarrow\chi\bar{\chi}$, is related to the mass of DM particles by\footnote{Here we are assuming that the produced DM particles in the annihilation $\gamma\gamma\rightarrow\chi\bar{\chi}$ are completely non-relativistic. Small corrections to this assumption do not change our result since all the limits presented in this paper are derived for large bins of $W_{\gamma\gamma}$.} $W_{\gamma\gamma}=2m_{\rm DM}$.

%%%%%%%%%            figure 2        %%%%%%%%%%%%%%
%%%%%%%%%%%%%%%%%%%%%%%%%%%%%%%%%
%%%%%%%%%%%%%%%%%%%%%%%%%%%%%%%%%%
%%%%%%%%%%%%%%%%%%%%%%%%%%%%%%%%%%

The cross section of $\gamma\gamma\rightarrow\chi\bar{\chi}$ can be measured by the {\it observation} of $pp+\slashed{E}$ events at the LHC overshooting the expected background; or the other way around, by the {\it non-observation} of any excess over the expected background, the cross section of $\gamma\gamma\rightarrow\chi\bar{\chi}$ can be constrained. The background events are the only limiting factor in this search and in this section we elaborate on it. 

%%%%%%%%%%%%%%%%%%%%%%%%%%%%%%%%
%%%%%%%%%%%%%%%%%%%%%%%%%%%%%%%%
\subsection{\label{sec:SMbkg}Backgrounds from the $\gamma\gamma\rightarrow$~SM processes}
%%%%%%%%%%%%%%%%%%%%%%%%%%%%%%%%
%%%%%%%%%%%%%%%%%%%%%%%%%%%%%%%%

One of the sources of background processes to the signal $pp+\slashed{E}$ is the whole processes $\gamma\gamma\rightarrow$~SM with final particles not passing through the central detectors. The LHC multipurpose detectors, ATLAS and CMS, have a pseudo-rapidity coverage range of $ |\eta| < 2.5$ where all sub-detectors (tracking system, electromagnetic and hadronic calorimeters, and the muon system) are available. All the SM processes with anything in the final state flying in the pseudo-rapidity range of $|\eta| > 2.5$ contribute to the background of our analysis. These background processes can be classified as follows:
\begin{itemize}
\item $l^{+}l^{-}$
\begin{eqnarray}
pp\rightarrow p~+ &\gamma\gamma&+~p, \nonumber \\
&\gamma\gamma&\rightarrow l^{+}l^{-}, ~ {\rm where}~l=e,\mu,\tau;~\text{with}~|\eta_{l}| > 2.5. \nonumber 
\end{eqnarray}
\item $q \bar{q}$
\begin{eqnarray}
pp\rightarrow p~+ &\gamma\gamma&+~p, \nonumber \\
&\gamma\gamma&\rightarrow q\bar{q}, ~\text{where}~q=u,d,c,s,b;~\text{with}~|\eta_{q}| > 2.5. \nonumber 
\end{eqnarray}
\item $W^{+}W^{-}$
\begin{eqnarray}
pp\rightarrow p~+ &\gamma\gamma&+~p, \nonumber \\
&\gamma\gamma&\rightarrow W^{+}W^{-}, ~\text{with}~W\rightarrow l\nu_{l},q\bar{q};~\text{with}~|\eta_{l,q}| > 2.5. \nonumber 
\end{eqnarray}
\end{itemize}
The cross section of all the above background processes have been calculated for various bins of the two-photon invariant mass ranging from 100~GeV to 1.2~TeV (see Table~\ref{tab:xsection}). The reported values have been calculated using the Monte Carlo event generators {\sc FPMC}~\cite{Boonekamp:2011ky} and {\sc MadGraph 5}~\cite{Alwall:2014hca} which simulate the two-photon exclusive production with the forward detector acceptance $0.0015 < \xi < 0.15$. The values of the  cross sections have been cross-checked with these two generators and reasonable agreement has been found. Basically, these generators calculate the cross sections by convoluting the probability of photon emission from the protons based on EPA with the photon-photon fusion cross section. The reported values are the cross sections after requiring all final state particles to be out of pseudo-rapidity range of the central detectors, i.e. $|\eta| > 2.5$ while a very low threshold of 5~GeV on the transverse momentum of the final state particles has been implemented.

%%%%%%%%%%%    Table 1    %%%%%%%%%%%%%%%%%
%%%%%%%%%%%%%%%%%%%%%%%%%%%%%%%%%%
\begin{footnotesize}
\begin{table}[h!]
\caption{\label{tab:xsection}The cross section of background processes $\gamma\gamma\rightarrow l^{+}l^{-}$, $\gamma\gamma\rightarrow q\bar{q}$ and $\gamma\gamma\rightarrow W^{+}W^{-}$ in various bins of the invariant mass of the final state particles. The reported values are the cross sections after requiring all the final state particles to be out of the central detectors pseudorapidity range, i.e. $|\eta| > 2.5$. All the cross sections are in fb.}
\begin{center}
\begin{tabular}{|c|c|c|c|}
\hline
Invariant mass of final state [GeV] & $l^{+}l^{-}$ & $q\bar{q}$ & $W^{+}W^{-}$ \\
\hline
\hline
$[100,300]$ & 1.83 & 0.70 & 0.43 \\
\hline
$[300,500]$ & 0.16 & 0.09 & 0.072 \\
\hline
$[500,700]$ & 0.05 & 0.02 & 0.02 \\
\hline
$[700,900]$ & 0.017 & 0.006 & 0.019 \\
\hline
$[900,1200]$ & 0.0025 & 0.002 & 0.01 \\
\hline
\end{tabular}
\end{center}
\end{table}
\end{footnotesize}
%%%%%%%%%%%%%%%%%%%%%%%%%%%%%%%%%%
%%%%%%%%%%%%%%%%%%%%%%%%%%%%%%%%%%

In addition to the above processes, another source of background needs to be considered which is the background contribution arising from the limited resolution of jet energy measurement in the hadron calorimeters of ATLAS and CMS detectors. The uncertainty on jet energy measurement depends on the jet transverse momentum ($p_{\rm T,j}$) and pseudo-rapidity. This uncertainty peaks at low $p_{\rm T,j}$ and large pseudo-rapidity and decreases with increasing the $p_{\rm T,j}$ in small pseudo-rapidity region. As a result, processes containing only low-$p_{\rm T,j}$ jets are usually discarded due to the large uncertainty and noise on low energy measurements. Therefore, the SM processes with two intact protons in the final state and quarks (jets) with transverse momentum smaller than $30$~GeV and pseudo-rapidity $|\eta| < 2.5$ are considered as a source of background in this study. The production rate for this background above the invariant mass cut of $100$~GeV is found to be $\sim 1$~fb.
 
Double Pomeron Exchange (DPE)~\cite{dpe1,dpe2} production of $W^+W^-$, dilepton and di-jet are additional sources of background processes. The cross sections of these backgrounds have been calculated using {\sc FPMC}~\cite{Boonekamp:2011ky}, and found to be quite negligible in the large invariant mass region with $|\eta| > 2.5$ and very low threshold on the transverse momentum of the final state particles~\cite{Boonekamp:2011ky,wwa}. As a result, these processes are not considered in this study.

%%%%%%%%%%%%%%%%%%%%%%%%%%%%%%%%
%%%%%%%%%%%%%%%%%%%%%%%%%%%%%%%%
\subsection{\label{sec:pile}Pile-up events}
%%%%%%%%%%%%%%%%%%%%%%%%%%%%%%%%
%%%%%%%%%%%%%%%%%%%%%%%%%%%%%%%%

At the LHC during the bunch crossing more than one proton-proton interaction can happen, the so-called pile-up interactions. The pile-up events can deteriorate the signal observation in two ways: {\it i}) a pile-up event can hinder the observation of signal. Since the signal signature of interest is the lack of activity in the central detector, occurrence of a simultaneous interaction between the protons in the same bunch crossing will mask this signature. In principle it is possible to reject the pile-up events by reconstructing the vertex of interaction. However, for the signal configuration considered in this paper there is no vertex reconstruction by the central detector; although it is possible to use the proton tagging in forward detectors for this purpose. The viability of vertex reconstruction (and so the ejection of pile-up events) requires high-precision measurement of the proton time of flight ($\sim10$~ps resolution in the measurement of the relative arrival time of protons to the forward detectors~\cite{Albrow:2008pn}) and fast communication between the forward and central detectors in order to use this information at trigger level. Although these are achievable (and requires more detailed studies), at the moment a practical way to overcome the pile-up events issue is to limit the data-taking to low instantaneous luminosity periods of the LHC, that is $\sim10^{33}~{\rm cm}^{-2}~{\rm s}^{-1}$ (see~\cite{Belotsky:2004ex} for a more detailed estimation of this background). {\it ii}) The pile-up events can mimic the signature of the signal. When the pile-up interactions take place through the hard non-diffractive processes, protons from the pile-up interactions within the acceptance of the forward detector can mimic the signal and so are backgrounds to our signal. In order to estimate this type of background, one should calculate the probability of observing such (accumulated) events in the forward detectors, which depends on the beam optic and the distance between forward detector and beam center and is $\sim (0.01-0.02)\%$ depending on the specifications of the forward detector and the beam properties~\cite{Trzebinski:2015bra}. The main contribution of this pile-up background to our signal comes from $ZZ$ or $W^+W^-$ production, categorized as follows:
\begin{itemize}
\item{$ZZ \rightarrow 4\nu$},
\item{$ZZ \rightarrow 2\nu f\bar{f}$}, where the fermion $f$ is out of detector's rapidity acceptance,
\item{$W^+W^- \rightarrow 2l2\nu$}, where the lepton $l$ is out of detector's rapidity acceptance,
\item{$W^+W^- \rightarrow l\nu q q'$}, where the lepton $l$ and quarks ($q,q'$) are out of detector's rapidity acceptance.
\end{itemize} 
Considering the probability of $0.01\%$ for observing a double tagged event with the above hard non-diffractive processes, the cross section of $ZZ$ and $W^+W^-$ processes are $0.06$~fb and $0.13$~fb, respectively. 

%%%%%%%%%%%%%%%%%%%%%%%%%%%%%%%%
%%%%%%%%%%%%%%%%%%%%%%%%%%%%%%%%
\subsection{\label{sec:brem}Bremsstrahlung of the beam protons}
%%%%%%%%%%%%%%%%%%%%%%%%%%%%%%%%
%%%%%%%%%%%%%%%%%%%%%%%%%%%%%%%%

\begin{figure}[h!]
\centering
\includegraphics[width=.7\textwidth]{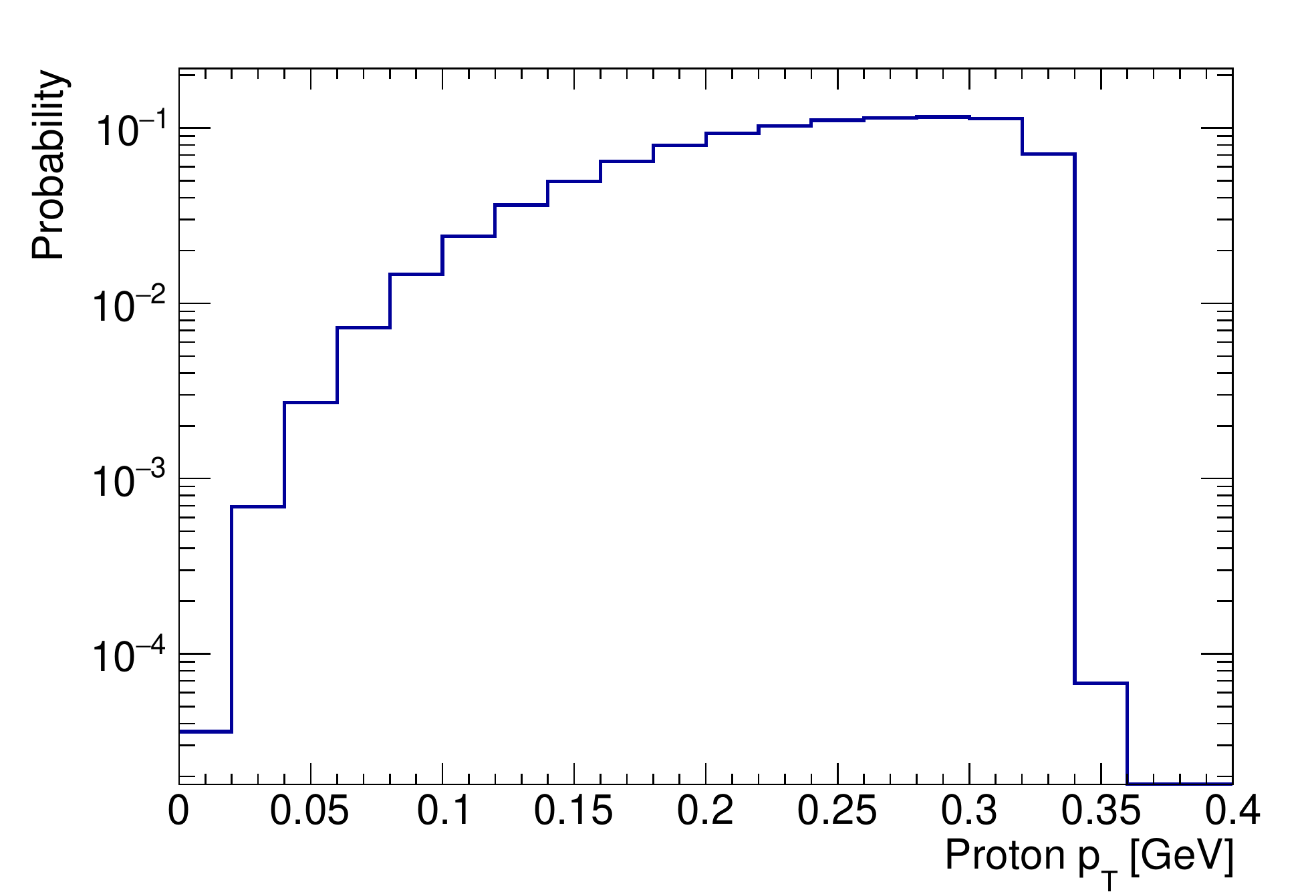}
\caption{\label{fig:brem} The distribution of the $p_{\rm T}$ of protons after bremsstrahlung emission, for proton beam with $\sqrt{s}=13$~TeV.}
\end{figure}

The bremsstrahlung emission from the beam protons, that is the process $pp\rightarrow pp\gamma\gamma$ due to QED radiation, have the same signature as our signal. In this process, each of the protons emit a $\gamma$ and lose part of its energy; the energy-degraded protons will pass through the forward detectors and mimic the signature. The cross section of the simultaneous bremsstrahlung radiation from both protons is $\sim$~pb (see~\cite{Belotsky:2004ex}) and in fact this background is the most severe one in our analysis. However, this background can be rejected in two ways: {\it i}) the bremsstrahlung emission process can be identified by observing the emitted photons which are strongly collimated in the beam direction (with the average emission angle $\sim m_p/E_p\simeq10^{-4}$~rad at $\sqrt{s}=13$~TeV). Tagging these photons requires a detector sensitive to photons in the vey forward direction, a task that can be accomplished by the Zero Degree Calorimeter (ZDC) detectors~\cite{zdc}. The electromagnetic module of the ZDCs are dedicated detectors for observation of neutral particles in the very forward direction covering the range $|\eta|>8.5$. The spatial extension of the ZDC detector covers the majority of bremsstrahlung photons emitted from protons with energy $6.5$~TeV, although not all of them (see Figure~3 of~\cite{Czekierda:2016fgz}). Thus, a large part of this background can be vetoed by using ZDC, a task that by a detector with efficiency $\sim97\%$ can be accomplished~\cite{Belotsky:2004ex}. {\it ii}) The other way of identifying the bremsstrahlung emission is the measurement of the transverse momenta of protons, $p_{\rm T}$, in the forward detectors after the emission of bremsstrahlung photons. Figure~\ref{fig:brem} shows the $p_{\rm T}$-distribution of protons after photon emission, generated by the GenEx Monte Carlo event generator~\cite{Kycia:2014hea}. As can be seen application of a cut $p_{\rm T}>0.4$~GeV on the transverse momenta of protons in the forward detectors can completely reject this background. 

%%%%%%%%%%%   Figure 3   %%%%%%%%%%%%%%
%%%%%%%%%%%%%%%%%%%%%%%%%%%%%%%
%%%%%%%%%%%%%%%%%%%%%%%%%%%%%%%
%%%%%%%%%%%%%%%%%%%%%%%%%%%%%%%

The most efficient rejection of the background events induced by the bremsstrahlung of protons is a combination of methods ({\it i}) and ({\it ii}). A cut $p_{\rm T}>0.4$~GeV will strongly suppress the effective photon-photon luminosity (see Figure~\ref{fig:lum}) which consequently diminish the potential of LHC in constraining the cross section of $\gamma\gamma\rightarrow\chi\bar{\chi}$; but, however, using the ZDC detector enables us to relax the cut on $p_{\rm T}$. Determination of the exact value of the $p_{\rm T}^{\rm cut}$ requires a detailed simulation of the ZDC detector which is out of the scope of this paper. Alternatively, in the next section we will present the potential of the LHC in constraining the process $\gamma\gamma\rightarrow\chi\bar{\chi}$ for various values of the $p_{\rm T}^{\rm cut}$.

Similar processes with pion production, such as $pp\rightarrow pp\pi^0$~\cite{Lebiedowicz:2013vya} and $pp\rightarrow pp\pi^+\pi^-$~\cite{Lebiedowicz:2009pj,Staszewski:2011bg} also seems to contribute to the background events. However, in these processes the energy losses of the protons are very small~\cite{Staszewski:2011bg} and for the central invariant masses we are interested in this paper, the contribution of these processes are completely negligible. Another process that can contribute to the background is the double-diffractive dissociation~\cite{Khoze:2000wk}. In the double-diffractive dissociation although both the protons dissociate, there is a non-negligible probability of finding protons in the dissociation products with lower energies that pass through the forward detectors. However, in this process always at least 4 more pions will be produced and it can be shown that at least one these pions will pass through the central detectors~\cite{Belotsky:2004ex}. So, this process can be easily vetoed by the central detector activity. 

%%%%%%%%%%%%%%%%%%%%%%%%%%%%%%%%
%%%%%%%%%%%%%%%%%%%%%%%%%%%%%%%%
\subsection{\label{sec:sen}Sensitivity of the LHC}
%%%%%%%%%%%%%%%%%%%%%%%%%%%%%%%%
%%%%%%%%%%%%%%%%%%%%%%%%%%%%%%%%

For a given integrated luminosity of the LHC, $\mathcal{L}$, and background cross section $\sigma_{\rm bg}$ (that is the sum of the all the processes discussed in the subsections~\ref{sec:SMbkg} and \ref{sec:pile}), the number of expected background events is $n_{\rm bg} = \epsilon \times \mathcal{L} \times \sigma_{\rm bg}$, where the overall efficiency $\epsilon = \epsilon_{\rm pt} \times \epsilon_{\rm sp}$ takes into account the proton tagging efficiency in the forward detectors ($\epsilon_{\rm pt}$) and proton survival probability ($\epsilon_{\rm sp}$), where the latter is the probability of additional soft gluon exchange between the incoming protons. For the elastic photon emission processes $\epsilon_{\rm sp}\simeq 0.9$~\footnote{In a more realistic analysis the dependence of $\epsilon_{\rm sp}$ on the two-photon invariant mass should be considered~\cite{sp3}.}~\cite{sp1,sp2}. For the proton tagging efficiency we assume $\epsilon_{\rm pt}=1$. To account for the uncertainties that arise from our assumptions, and also other possible sources of uncertainties, an overall (conservative) uncertainty of $10\%$ is assigned to the $\epsilon$ value in the extraction of limits. Also, an uncertainty of $2.5\%$ is considered on the integrated luminosity~\cite{lumi1}.

\begin{figure}[h!]
\centering
\includegraphics[width=.7\textwidth]{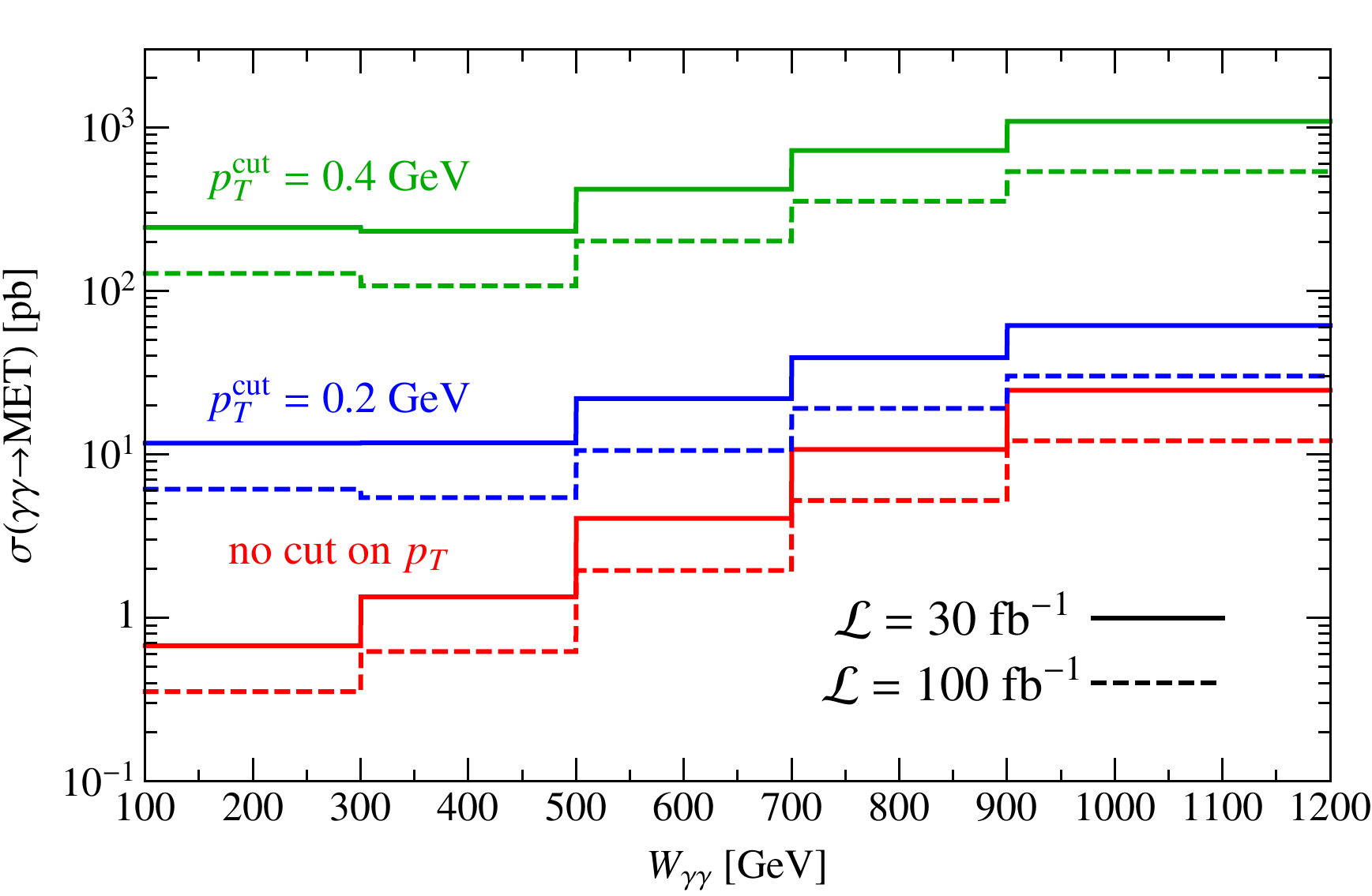}
\caption{\label{fig:lim} The expected upper limit on the cross section of $pp\rightarrow p+\gamma\gamma+p\rightarrow pp+\slashed{E}$, at $95\%$~C.L., in terms of the two-photon invariant mass. The solid and dashed curves correspond to the LHC luminosity $\mathcal{L}=30$~fb$^{-1}$ and $\mathcal{L}=100$~fb$^{-1}$, respectively. The red, blue and green curves show the expected sensitivity for no cut on $p_{\rm T}$, $p_{\rm T}^{\rm cut}=0.2$~GeV and $p_{\rm T}^{\rm cut}=0.4$~GeV, respectively.}
\end{figure}

The sensitivity of the LHC (that is the projected upper limit that can be set by the LHC) to the process $p p\rightarrow p+\gamma \gamma +p\rightarrow p+\chi\bar{\chi}+p$ can be obtained by using the Poisson statistics. To estimate the sensitivity of LHC, we assume that the observed number of events at the LHC, $n_{\rm obs}$, will be consistent with the expected number of background such that $n_{\rm obs} = \lceil n_{\rm bg} \rceil$ where $\lceil x \rceil$ denotes the ceiling function, that is the smallest integer greater than or equal to $x$~\footnote{Relaxing the condition $n_{\rm obs} = \lceil n_{\rm bg} \rceil$ to $n_{\rm obs} \sim \mathcal{O}\left(\lceil n_{\rm bg} \rceil\right)$ will change the reported results by a factor of few. Obviously, the case $n_{\rm obs} \gg n_{\rm bg}$ corresponds to a {\it discovery} of new physics which is responsible for this excess of $pp\rightarrow p+\gamma\gamma+p\rightarrow pp+\slashed{E}$ events; and so instead of calculating {\it upper limit}, which we do in this paper, one should calculate the projected {\it significance} of discovery.}. The upper limit on the cross section of $\gamma\gamma\rightarrow\chi\bar{\chi}$ can be derived by requiring $n<n_{\rm limit}$ where $n$ is the induced number of events from $\gamma\gamma\rightarrow\chi\bar{\chi}$ reaction (which is $\mathcal{L}_{\gamma\gamma} \times\sigma\left(\gamma\gamma\rightarrow\chi\bar{\chi}\right)$), and $n_{\rm limit}$ is extracted from the following equation\footnote{Equivalently, the $n_{\rm limit}$ can be extracted from the following equation (at $q\%$ C.L.):
\begin{equation}\label{eq:stat2} 
1-\frac{q}{100} = \frac{\displaystyle\sum_{m=0}^{n_{\rm obs}} \frac{\left( n_{\rm limit} + n_{\rm bg}\right)^m}{m!}}{\displaystyle\sum_{m=0}^{n_{\rm obs}} \frac{\left( n_{\rm bg}\right)^m}{m!}}~e^{-n_{\rm limit}}~. 
\end{equation}
The value of $n_{\rm limit}$ extracted from Eqs.~(\ref{eq:stat1}) and (\ref{eq:stat2}), dubbed respectively the Bayesian and frequentist approaches, are the same.} at a given confidence level of $q\%$:
\begin{eqnarray}\label{eq:stat1}
\frac{q}{100} = \frac{\displaystyle\int_0^{n_{\rm limit}} L(n_{\rm obs},N)~{\rm d}N}{\displaystyle\int_0^\infty  L(n_{\rm obs},N)~{\rm d}N}~,
\end{eqnarray}
where
\begin{equation}
L\left( n_{\rm obs},N \right) = \frac{\left( N + n_{\rm bg}\right)^{n_{\rm obs}}}{n_{\rm obs}!} ~e^{-\left( N+n_{\rm bg}\right)}~.
\end{equation}
In Table~\ref{tab:Sfix} we report the sensitivity of LHC, at $95\%$ C.L., to $\sigma(\gamma\gamma\rightarrow\chi\bar{\chi})$ in various bins of the invariant mass of two-photon system, for two integrated luminosity values of $\mathcal{L}=30$ and $100$~fb$^{-1}$. All the limits reported in Table~\ref{tab:Sfix} are calculated without any cut on the $p_{\rm T}$. As we discussed in subsection~\ref{sec:brem}, no cut on the $p_{\rm T}$ of protons means we are assuming that all the background events from the bremsstrahlung process can be rejected by the ZDC detector. Since achieving this goal seems too optimistic (though it is not impractical and requires a more detailed study of the ZDC detector) we have calculated also the sensitivity of LHC with the implementation of cuts $p_{\rm T}^{\rm cut}=0.2$~GeV and $p_{\rm T}^{\rm cut}=0.4$~GeV. Figure~\ref{fig:lim} shows these sensitivities as function of $W_{\gamma\gamma}$. In this figure, the solid (dashed) curves correspond to the LHC luminosity $\mathcal{L}=30$ ($100$)~fb$^{-1}$. As can be seen, applying the $p_{\rm T}^{\rm cut}=0.2$~GeV degrades the sensitivity in low $W_{\gamma\gamma}$ values by almost one order of magnitude, while for the high $W_{\gamma\gamma}$ values the sensitivity worsen by a factor of few. Increasing the cut to $p_{\rm T}^{\rm cut}=0.4$~GeV (which is a very pessimistic scenario) will worsen the sensitivity by $\sim2$ orders of magnitude for all the $W_{\gamma\gamma}$ values. 

For the large values of $W_{\gamma\gamma}$, the number of SM background events is almost zero while the pile-up events still contribute to the background. The sensitivity spoils for higher values of $W_{\gamma\gamma}$ since the photon luminosity drops rapidly (see Figure~\ref{fig:lum}).

%%%%%%%%%%%    Table 2    %%%%%%%%%%%%%%%%%
%%%%%%%%%%%%%%%%%%%%%%%%%%%%%%%%%%
\begin{footnotesize}
\begin{table}[t!]
\caption{\label{tab:Sfix}The sensitivity of the LHC to $\sigma(\gamma\gamma\rightarrow\chi\bar{\chi})$ ($\langle\sigma(\chi\bar{\chi}\rightarrow\gamma\gamma)v\rangle_{\gamma\gamma}$) in the unit pb (cm$^3$/s) for $\mathcal{L}=30$ and $100$~fb$^{-1}$. All the numbers are at $95\%$ C.L. All the limits are calculated by the assumption of no cut on the $p_{\rm T}$ of protons.}
\begin{center}
\begin{tabular}{|c|c|c|}
\hline
$W_{\gamma \gamma}$ [GeV] & \rm{Sensitivity to} $\sigma$ ($\langle\sigma v\rangle_{\gamma\gamma}$) at 30 fb$^{-1}$ & \rm{Sensitivity to} $\sigma$ ($\langle\sigma v\rangle_{\gamma\gamma}$) at 100 fb$^{-1}$\\
\hline
\hline
$[100,300]$ & $0.68$ ($2.03\times10^{-29}$) & $0.35$ ($1.06\times10^{-29}$)\\
\hline
$[300,500]$ & $1.34$ ($4.03\times10^{-29}$) & $6.21$ ($1.86\times10^{-29}$)\\
\hline
$[500,700]$ & $4.05$ ($1.22\times10^{-28}$) & $1.95$ ($5.84\times10^{-29}$)\\
\hline
$[700,900]$ & $10.7$ ($3.21\times10^{-28}$) & $5.22$ ($1.57\times10^{-28}$)\\
\hline
$[900,1200]$ & $24.6$ ($7.39\times10^{-28}$) & $12.1$ ($3.63\times10^{-28}$)\\
\hline
\end{tabular}
\end{center}
\end{table}
\end{footnotesize}
%%%%%%%%%%%%%%%%%%%%%%%%%%%%%%%%%%
%%%%%%%%%%%%%%%%%%%%%%%%%%%%%%%%%%

%%%%%%%%%%%   Figure 4   %%%%%%%%%%%%%%
%%%%%%%%%%%%%%%%%%%%%%%%%%%%%%%
%%%%%%%%%%%%%%%%%%%%%%%%%%%%%%%
%%%%%%%%%%%%%%%%%%%%%%%%%%%%%%%

\begin{figure}[t!]
\centering
\includegraphics[width=.8\textwidth]{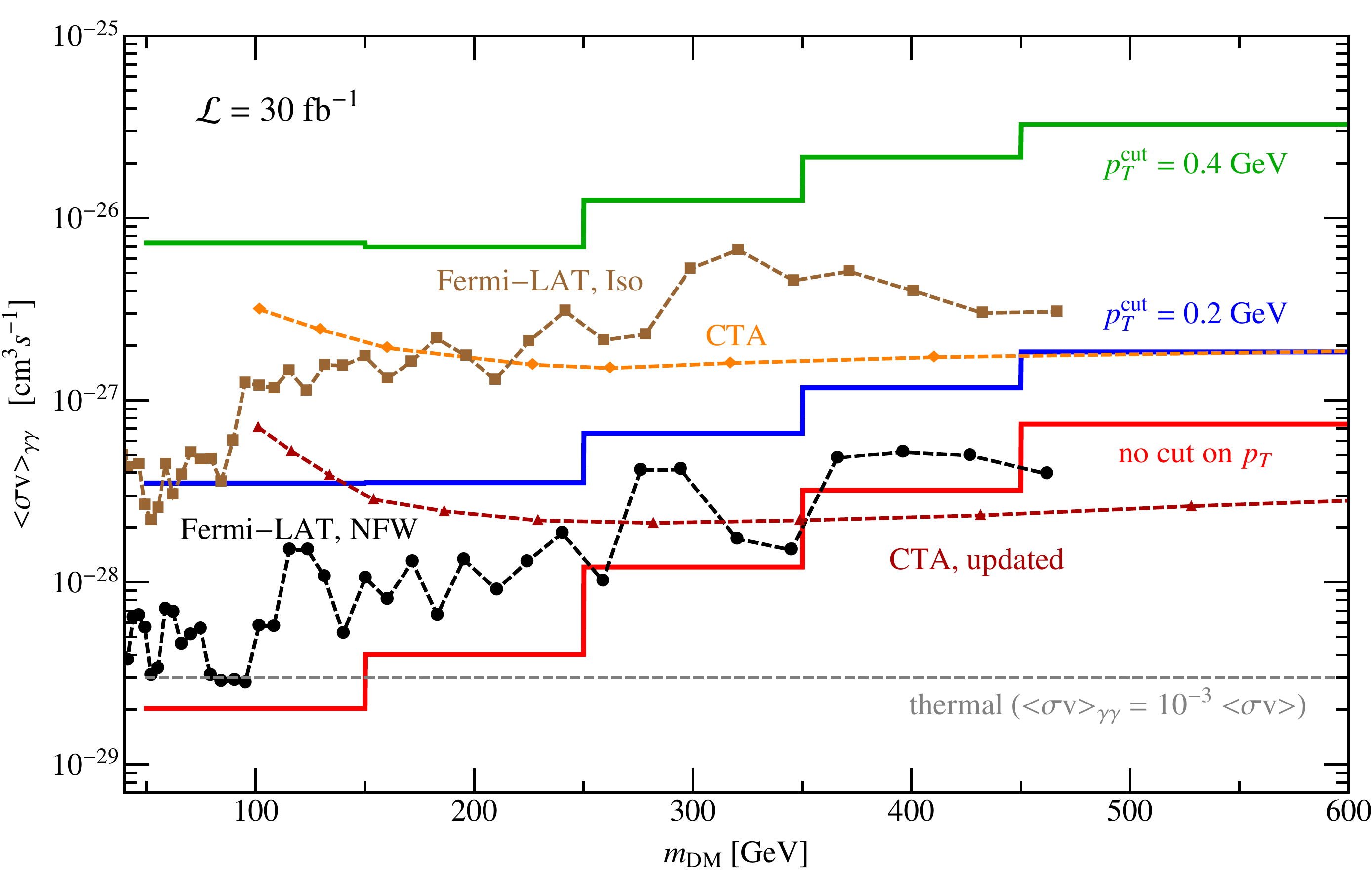}
\caption{\label{fig:lim2} The sensitivity of the LHC, at $95\%$~C.L. and $\mathcal{L}=30$~fb$^{-1}$, to the cross section of dark matter annihilation to monochromatic gamma rays, $\langle\sigma v\rangle_{\gamma\gamma}\equiv\langle\sigma(\chi\bar{\chi}\rightarrow\gamma\gamma)v\rangle$, for dark matter masses $m_{\rm DM}\sim(50-600)$~GeV and for three different assumptions on $p_{\rm T}^{\rm cut}$ (we have assumed $v\simeq10^{-3}c$ in the calculation of LHC sensitivity). The red curve shows the sensitivity of the LHC without applying any cut on the transverse momentum of forward protons; while the blue and green curves show the sensitivity by applying the cuts $p_{\rm T}^{\rm cut} = 0.2$~GeV and $p_{\rm T}^{\rm cut} = 0.4$~GeV, respectively. The black and brown curves show the upper limits on $\langle\sigma v\rangle_{\gamma\gamma}$ from Fermi-LAT experiment assuming cuspy NFW and isothermal dark matter profile respectively~\cite{Ackermann:2015lka}. The orange and dark-red curves show the sensitivity of CTA, respectively for nominal~\cite{Bernlohr:2012we} and updated~\cite{updatedCTA} performances (see~\cite{Ibarra:2015tya}). The dashed gray line shows the expected $\langle\sigma v\rangle_{\gamma\gamma}$ for a thermal dark matter scenario with $\langle\sigma v\rangle_{\gamma\gamma}/\langle\sigma v\rangle=10^{-3}$.}
\end{figure}

\begin{figure}[t!]
\centering
\includegraphics[width=.8\textwidth]{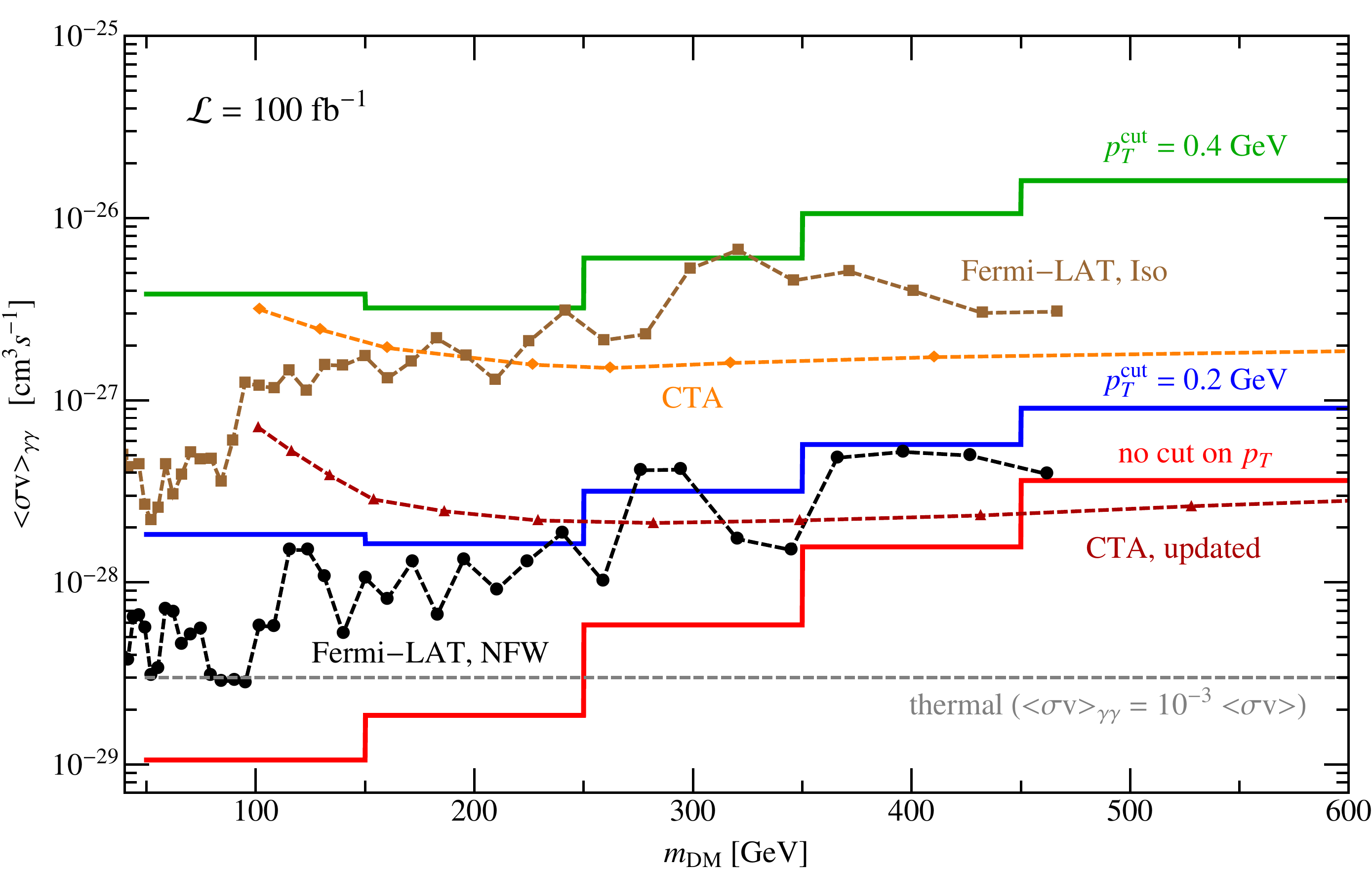}
\caption{\label{fig:lim100} The same as Figure~\ref{fig:lim2} but for $\mathcal{L}=100$~fb$^{-1}$.}
\end{figure}

Due to the crossing symmetry, the sensitivity presented in Figure~\ref{fig:lim} can be translated to the sensitivity of the LHC to the monochromatic gamma-ray production in dark matter annihilation. In order to compare with the current limits on $\sigma(\chi\bar{\chi}\rightarrow\gamma\gamma)$ which comes from indirect searches, we show in Figure~\ref{fig:lim2} the sensitivity of the LHC, at $95\%$ C.L. and $\mathcal{L}=30$~fb$^{-1}$ and for three different assumptions on $p_{\rm T}^{\rm cut}$, to the $\langle\sigma v\rangle_{\gamma\gamma}\equiv\langle\sigma(\chi\bar{\chi}\rightarrow\gamma\gamma)v\rangle$ by assuming $v\simeq10^{-3}c$ (as the average velocity of non-relativistic dark matter particles at the present time). The corresponding sensitivity in bins of $W_{\gamma\gamma}$ is reported in Table~\ref{tab:Sfix}. By the crossing symmetry, the invariant mass of the two-photon system $W_{\gamma\gamma}$ corresponds to $2m_{\rm DM}$ (assuming the production of non-relativistic DM particles in the $\gamma\gamma$ annihilation). The black and brown curves in Figure~\ref{fig:lim2} show the upper limits on $\langle\sigma v\rangle_{\gamma\gamma}$ from Fermi-LAT experiment in the search for spectral line from direct annihilation of dark matter to gamma-ray in the Milky Way, with {\sc Pass 8} data, assuming cuspy NFW and isothermal dark matter profile respectively~\cite{Ackermann:2015lka}. As in all the indirect searches, the limit of Fermi-LAT depends on the assumed dark matter profile, and the two curves of Fermi-LAT in Figure~\ref{fig:lim2} brackets this uncertainty. The Fermi-LAT limits extend up to $\sim500$~GeV that is the highest energy of gamma-rays detectable by Fermi-LAT instrument. For higher masses the H.E.S.S. limit applies which is obtained by looking for line-like spectral features from Galactic center region and assuming Einasto dark matter profile~\cite{Abramowski:2013ax}. The H.E.S.S. limit is not shown in the figure (since just the lower tail of the limit fit in the figure) but it is at the same ballpark of the limit from Fermi-LAT. The orange and dark-red curves in Figure~\ref{fig:lim2} show the sensitivity of near-future Cherenkov Telescope Array (CTA), respectively for nominal~\cite{Bernlohr:2012we} and updated~\cite{updatedCTA} performances (for the details see~\cite{Ibarra:2015tya} where the limits has been taken from). The horizontal dashed gray curve in Figure~\ref{fig:lim2} shows the expected $\langle\sigma v\rangle_{\gamma\gamma}$ for a dark matter scenario with total annihilation cross section $\langle\sigma v\rangle=3\times10^{-26}~{\rm cm}^3~{\rm s}^{-1}$ (which meets the requirement of thermal freeze-out mechanism to provide the right present time abundance of dark matter particles) and assuming $\langle\sigma v\rangle_{\gamma\gamma}=10^{-3} \langle\sigma v\rangle$. The ratio of $\langle\sigma v\rangle_{\gamma\gamma}/\langle\sigma v\rangle$ is model-dependent and falling usually in the range $10^{-4}-10^{-1}$, while smaller ratios can arise in many models. As can be seen from the Figure~\ref{fig:lim2}, without applying any cut on the $p_{\rm T}$, the LHC can exclude models of dark matter which are based on the freeze-out mechanism and predict the $\langle\sigma v\rangle_{\gamma\gamma}/\langle\sigma v\rangle \gtrsim 10^{-3}$ (for a dark matter particle mass in the range $\sim (50-300)$~GeV). By implementing $p_{\rm T}^{\rm cut}=0.2$~GeV, the LHC is sensitive to $\langle\sigma v\rangle_{\gamma\gamma}$ roughly (better) than the CTA sensitivity for DM masses up to 600~GeV. Increasing the $p_{\rm T}^{\rm cut}$ to $0.4$~GeV will degrade the LHC sensitivity further to the level $\langle\sigma v\rangle_{\gamma\gamma}\sim 10^{-26}~{\rm cm}^{3}~{\rm s}^{-1}$. Figure~\ref{fig:lim100} is similar to Figure~\ref{fig:lim2} except for $\mathcal{L}=100$~fb$^{-1}$. Increasing the luminosity $\mathcal{L}$ is just increasing the statistics, which can be seen by comparing the Figures~\ref{fig:lim2} and \ref{fig:lim100}.     

%%%%%%%%%%%   Figure 5   %%%%%%%%%%%%%%
%%%%%%%%%%%%%%%%%%%%%%%%%%%%%%%
%%%%%%%%%%%%%%%%%%%%%%%%%%%%%%%
%%%%%%%%%%%%%%%%%%%%%%%%%%%%%%%

%%%%%%%%%%%   Figure 6   %%%%%%%%%%%%%%
%%%%%%%%%%%%%%%%%%%%%%%%%%%%%%%
%%%%%%%%%%%%%%%%%%%%%%%%%%%%%%%
%%%%%%%%%%%%%%%%%%%%%%%%%%%%%%%

The DM mass range shown in Figures~\ref{fig:lim2} and \ref{fig:lim100}, that is $m_{\rm DM}\sim(50-600)$~GeV, is dictated by the background processes and the effective luminosity of the photon emission from the protons. For $m_{\rm DM}\lesssim50$~GeV, the cross section of SM backgrounds and also the uncertainties grows very rapidly such that deriving a reliable bound on the $\langle\sigma v\rangle_{\gamma\gamma}$ will be practically unfeasible. On the other hand, for $m_{\rm DM}\gtrsim600$~GeV the SM backgrounds are almost zero (tough the pile-up events exist), but since the flux of photons emitted from protons decreases by increasing the energy (see Figure~\ref{fig:lum}), the sensitivity for $m_{\rm DM}\gtrsim600$~GeV deteriorate quickly.

%%%%%%%%%%%%%%%%%%%%%%%%%%%%%%
\section{Conclusions}\label{sec:Conclusions}
%%%%%%%%%%%%%%%%%%%%%%%%%%%%%%

The main drawback in the orthodox searches for dark matter at colliders, such as the LHC, is the lack of knowledge whether the scattering between colliding particles occurred or not: since the produced dark matter in the scattering of colliding particles (protons in the case of LHC) leave the detector without any trace, that is lack of any registered activity in the detectors, the signature for the annihilation of colliding particles into dark matter particles is the same as the case where particles in the colliding beams simply pass on each other without any scattering. The price one need to pay for the remedy, is to look for higher order scattering amplitudes where the dark matter particles are produced in the scattering of colliding particles accompanied by some visible activities such as hadronic state, photon or $Z$ production.    

However, there is an opportunity in the near future to look for the lowest order dark matter production at the LHC in the following circumstance: with the installation of forward detectors at the CMS and ATLAS, it is possible to tag the outgoing protons from an {\it elastic} scattering mediated via the emitted photons from incoming protons that leaves the protons intact. The observation of scattered protons in the forward detectors provides information on the occurrence of scattering (measurement of the energy and transverse momenta of protons also reveals the energy of mediated photons); and so, lack of activity in the central detectors in coincidence with the registered intact protons in the forward detectors can be a signal of dark matter production in the scattering. In this circumstance, the dark matter production is through the photon-photon fusion ($\gamma\gamma\rightarrow\chi\bar{\chi}$), which is basically the inverse process of dark matter annihilation to monochromatic gamma rays ($\chi\bar{\chi}\rightarrow\gamma\gamma$), a process under investigation in indirect dark matter searches. Thus, according to the crossing symmetry, the LHC (with the implemented forward detectors) can constrain $\sigma(\chi\bar{\chi}\rightarrow\gamma\gamma)$ and in fact a limit competitive to the current and near future indirect dark matter searches, in the range $m_{\rm DM}\sim(50-600)$~GeV, can be obtained. 

We have studied in detail the sensitivity of LHC to $\sigma(\chi\bar{\chi}\rightarrow\gamma\gamma)$ (or equivalently, $\langle \sigma v\rangle_{\gamma\gamma}\equiv\langle\sigma(\chi\bar{\chi}\rightarrow\gamma\gamma)v\rangle$). The main limiting factor for the LHC is the background processes from several sources including the SM induced backgrounds, pile-up events and the bremsstrahlung of the incoming protons. The irreducible SM induced background, with the largest contribution coming from $\gamma\gamma\rightarrow l^+ l^-$ where the produced leptons do not pass through the central detector, has been calculated and taken into account. The pile-up events enforce low-luminosity periods of data-taking; although these background can be vetoed also by using the information of forward detectors in the vertex reconstruction. The bremsstrahlung process, the severest background, can be rejected by the ZDC detector and/or application of cut on the transverse momenta of scattered protons in the forward detectors, though the most efficient rejection would be a combination of the both methods. Considering all these backgrounds (and assuming ZDC rejection of bremsstrahlung background), we have shown that in the DM mass range $m_{\rm DM}\sim(50-600)$~GeV, the LHC has sensitivity to $\langle \sigma v\rangle_{\gamma\gamma}\sim (10^{-29}-10^{-27})~{\rm cm}^3~{\rm s}^{-1}$, which is comparable to the existing limits from Fermi-LAT and the projected sensitivity of CTA. Application of the $p_{\rm T}^{\rm cut}$ will worsen the sensitivity by $\sim$ one order of magnitude. Obviously, the expected value of $\langle \sigma v\rangle_{\gamma\gamma}$ for DM is model-dependent; but, however, many models predict a value $\left(10^{-4}-10^{-1}\right)\langle \sigma v\rangle$, where $\langle \sigma v\rangle=3\times10^{-26}~{\rm cm}^3~{\rm s}^{-1}$ is the total annihilation cross section of dark matter particles motivated by the thermal freeze-out mechanism. We have shown that the LHC can exclude any dark matter model with $m_{\rm DM}\sim(50-600)$~GeV that predict $\langle\sigma v\rangle_{\gamma\gamma}/\langle\sigma v\rangle \gtrsim \left(10^{-3}-10^{-2}\right)$. The advantage of this study is that the LHC limits are free from the astrophysical uncertainties (such as DM halo profile, precise DM density at the center of galaxies, etc).      

The projected sensitivity of the LHC has been presented for DM mass range $m_{\rm DM}\sim(50-600)$~GeV. The lower value of the range dictated from the SM induced backgrounds that increase rapidly for invariant masses $\lesssim100$~GeV. On the other hand, the upper value of the range comes from the rapid reduction of the effective photon-photon luminosity in proton scattering; the sensitivity will drop quickly for $m_{\rm DM}\gtrsim600$~GeV, due to the rapid drop in the photon luminosity.

Finally, we would like to emphasize that estimation of the feasibility of the proposal in this paper requires a detailed simulation of the forward and central detectors, ZDC capability in the rejection of bremsstrahlung background and achievement of necessary triggering level; which is clearly beyond the scope of the present paper and must be performed within the experimental collaborations. Hopefully this essay provokes these kind of studies.    

%%%%%%%%%%%%%%%%%%%%%%%%%
\section*{Acknowledgment}
A.~E. thanks A.~Belyaev, S.~Fichet and V.~Khoze for useful discussions; and thanks G.~Gersdorff and P.~Serpico for several discussions and valuable comments on the manuscript. We would like to thank the authors of GenEx package for sharing it with us, and especially thank J.~Chwastowski for clarifications. M.~M.~N. is grateful to Albert de Roeck for the useful comments. S.~K. would like to thank Iran National Science Foundation (INSF) for the financial support and the CERN Theory Division for the very nice hospitality. 
%%%%%%%%%%%%%%%%%%%%%%%%%

%%%%%%%%%%%%%%%%%%%%%%%

\end{document}